
\message
{JNL.TEX version 0.95 as of 5/13/90.  Using CM fonts.}

\catcode`@=11
\expandafter\ifx\csname inp@t\endcsname\relax\let\inp@t=\input
\def\input#1 {\expandafter\ifx\csname #1IsLoaded\endcsname\relax
\inp@t#1%
\expandafter\def\csname #1IsLoaded\endcsname{(#1 was previously loaded)}
\else\message{\csname #1IsLoaded\endcsname}\fi}\fi
\catcode`@=12

\font\twelverm=cmr12			\font\twelvei=cmmi12
\font\twelvesy=cmsy10 scaled 1200	\font\twelveex=cmex10 scaled 1200
\font\twelvebf=cmbx12			\font\twelvesl=cmsl12
\font\twelvett=cmtt12			\font\twelveit=cmti12
\font\twelvesc=cmcsc10 scaled 1200	\font\twelvesf=cmss12
                     
\skewchar\twelvei='177			\skewchar\twelvesy='60


\def\twelvepoint{\normalbaselineskip=12.4pt plus 0.1pt minus 0.1pt
  \abovedisplayskip 12.4pt plus 3pt minus 9pt
  \belowdisplayskip 12.4pt plus 3pt minus 9pt
  \abovedisplayshortskip 0pt plus 3pt
  \belowdisplayshortskip 7.2pt plus 3pt minus 4pt
  \smallskipamount=3.6pt plus1.2pt minus1.2pt
  \medskipamount=7.2pt plus2.4pt minus2.4pt
  \bigskipamount=14.4pt plus4.8pt minus4.8pt
  \def\rm{\fam0\twelverm}          \def\it{\fam\itfam\twelveit}%
  \def\sl{\fam\slfam\twelvesl}     \def\bf{\fam\bffam\twelvebf}%
  \def\mit{\fam 1}                 \def\cal{\fam 2}%
  \def\sc{\twelvesc}		   \def\tt{\twelvett}%
  \def\sf{\twelvesf}
  \textfont0=\twelverm   \scriptfont0=\tenrm   \scriptscriptfont0=\sevenrm
  \textfont1=\twelvei    \scriptfont1=\teni    \scriptscriptfont1=\seveni
  \textfont2=\twelvesy   \scriptfont2=\tensy   \scriptscriptfont2=\sevensy
  \textfont3=\twelveex   \scriptfont3=\twelveex\scriptscriptfont3=\twelveex
  \textfont\itfam=\twelveit
  \textfont\slfam=\twelvesl
  \textfont\bffam=\twelvebf \scriptfont\bffam=\tenbf
                            \scriptscriptfont\bffam=\sevenbf
  \normalbaselines\rm}


\mathchardef\alpha="710B
\mathchardef\beta="710C
\mathchardef\gamma="710D
\mathchardef\delta="710E
\mathchardef\epsilon="710F
\mathchardef\zeta="7110
\mathchardef\eta="7111
\mathchardef\theta="7112
\mathchardef\iota="7113
\mathchardef\kappa="7114
\mathchardef\lambda="7115
\mathchardef\mu="7116
\mathchardef\nu="7117
\mathchardef\xi="7118
\mathchardef\pi="7119
\mathchardef\rho="711A
\mathchardef\sigma="711B
\mathchardef\tau="711C
\mathchardef\phi="711E
\mathchardef\chi="711F
\mathchardef\psi="7120
\mathchardef\omega="7121
\mathchardef\varepsilon="7122
\mathchardef\vartheta="7123
\mathchardef\varpi="7124
\mathchardef\varrho="7125
\mathchardef\varsigma="7126
\mathchardef\varphi="7127


\def\beginlinemode{\endmode
  \begingroup\parskip=0pt \obeylines\def\\{\par}\def\endmode{\par\endgroup}}
\def\beginparmode{\endmode
  \begingroup \def\endmode{\par\endgroup}}
\let\endmode=\par
{\obeylines\gdef\
{}}
\def\singlespace{\baselineskip=\normalbaselineskip}

\def\oneandahalfspace{\baselineskip=\normalbaselineskip
  \multiply\baselineskip by 3 \divide\baselineskip by 2}
\def\doublespace{\baselineskip=\normalbaselineskip \multiply\baselineskip by 2}

\newcount\firstpageno
\firstpageno=2
\footline={\ifnum\pageno<\firstpageno{\hfil}\else{\hfil\twelverm\folio\hfil}\fi}
\def\toppageno{\global\footline={\hfil}\global\headline
  ={\ifnum\pageno<\firstpageno{\hfil}\else{\hfil\twelverm\folio\hfil}\fi}}
\let\rawfootnote=\footnote		
\def\footnote#1#2{{\rm\singlespace\parindent=0pt\parskip=0pt
  \rawfootnote{#1}{#2\hfill\vrule height 0pt depth 6pt width 0pt}}}
\def\raggedcenter{\leftskip=4em plus 12em \rightskip=\leftskip
  \parindent=0pt \parfillskip=0pt \spaceskip=.3333em \xspaceskip=.5em
  \pretolerance=9999 \tolerance=9999
  \hyphenpenalty=9999 \exhyphenpenalty=9999 }
\def\dateline{\rightline{\ifcase\month\or
  January\or February\or March\or April\or May\or June\or
  July\or August\or September\or October\or November\or December\fi
  \space\number\year}}
\def\received{\vskip 3pt plus 0.2fill
 \centerline{\sl (Received\space\ifcase\month\or
  January\or February\or March\or April\or May\or June\or
  July\or August\or September\or October\or November\or December\fi
  \qquad, \number\year)}}


\hsize=6.5truein
\hoffset=0pt
\vsize=8.9truein
\voffset=0pt
\parskip=\medskipamount
\def\\{\cr}
\twelvepoint		
\doublespace		
\overfullrule=0pt	




\def\title			
  {\null\vskip 3pt plus 0.2fill
   \beginlinemode \doublespace \raggedcenter \bf}

\def\author			
  {\vskip 3pt plus 0.2fill \beginlinemode
   \singlespace \raggedcenter\sc}

\def\affil			
  {\vskip 3pt plus 0.1fill \beginlinemode
   \oneandahalfspace \raggedcenter \sl}

\def\abstract			
  {\vskip 3pt plus 0.3fill \beginparmode
   \oneandahalfspace ABSTRACT: }

\def\endtitlepage		
  {\endpage			
   \body}
\let\endtopmatter=\endtitlepage

\def\body			
  {\beginparmode}		

\def\head#1{			
  \goodbreak\vskip 0.5truein	
  {\immediate\write16{#1}
   \raggedcenter \uppercase{#1}\par}
   \nobreak\vskip 0.25truein\nobreak}

\def\subhead#1{			
  \vskip 0.25truein		
  {\raggedcenter {#1} \par}
   \nobreak\vskip 0.25truein\nobreak}

\def\beginitems{
\par\medskip\bgroup\def\i##1 {\item{##1}}\def\ii##1 {\itemitem{##1}}
\leftskip=36pt\parskip=0pt}
\def\enditems{\par\egroup}

\def\beneathrel#1\under#2{\mathrel{\mathop{#2}\limits_{#1}}}

\def\refto#1{$^{#1}$}		

\def\references			
  {\head{References}		
   \beginparmode
   \frenchspacing \parindent=0pt \leftskip=1truecm
   \parskip=8pt plus 3pt \everypar{\hangindent=\parindent}}

\gdef\refis#1{\item{#1.\ }}			

\gdef\journal#1, #2, #3, 1#4#5#6{		
    {\sl #1~}{\bf #2}, #3 (1#4#5#6)}		

\def\prd{\journal Phys. Rev. D, }

\def\prl{\journal Phys. Rev. Lett., }

\def\endreferences{\body}

\def\figurecaptions		
  {\endpage
   \beginparmode
   \head{Figure Captions}
}

\def\endpage			
  {\vfill\eject}

\def\endpaper			
  {\endmode\vfill\supereject}

\def\endit
  {\endpaper\end}


\def\heading				
  {\vskip 0.5truein plus 0.1truein	
   \beginparmode \def\\{\par} \parskip=0pt \singlespace \raggedcenter}

\def\subheading				
  {\vskip 0.25truein plus 0.1truein	
   \beginlinemode \singlespace \parskip=0pt \def\\{\par}\raggedcenter}

\def\tag#1$${\eqno(#1)$$}

\def\align#1$${\eqalign{#1}$$}

\def\aligntag#1$${\gdef\tag##1\\{&(##1)\cr}\eqalignno{#1\\}$$
  \gdef\tag##1$${\eqno(##1)$$}}

\def\overset #1\to#2{{\mathop{#2}\limits^{#1}}}
\def\underset#1\to#2{{\let\next=#1\mathpalette\undersetpalette#2}}
\def\undersetpalette#1#2{\vtop{\baselineskip0pt
\ialign{$\mathsurround=0pt #1\hfil##\hfil$\crcr#2\crcr\next\crcr}}}


\def\ref#1{Ref.~#1}			
\def\Ref#1{Ref.~#1}			
\def\[#1]{[\cite{#1}]}
\def\cite#1{{#1}}
\def\(#1){(\call{#1})}
\def\call#1{{#1}}
\def\taghead#1{}
\def\frac#1#2{{#1 \over #2}}

\def\12{{1\over2}}
\def\eg{{\it e.g.,\ }}

\def\sla{\raise.15ex\hbox{$/$}\kern-.57em}
\def\leaderfill{\leaders\hbox to 1em{\hss.\hss}\hfill}
\def\twiddle{\lower.9ex\rlap{$\kern-.1em\scriptstyle\sim$}}
\def\bigtwiddle{\lower1.ex\rlap{$\sim$}}
\def\gtwid{\mathrel{\raise.3ex\hbox{$>$\kern-.75em\lower1ex\hbox{$\sim$}}}}
\def\ltwid{\mathrel{\raise.3ex\hbox{$<$\kern-.75em\lower1ex\hbox{$\sim$}}}}
\def\square{\kern1pt\vbox{\hrule height 1.2pt\hbox{\vrule width 1.2pt\hskip 3pt
   \vbox{\vskip 6pt}\hskip 3pt\vrule width 0.6pt}\hrule height 0.6pt}\kern1pt}
\def\tdot#1{\mathord{\mathop{#1}\limits^{\kern2pt\ldots}}}

\def\pmb#1{\setbox0=\hbox{#1}%
  \kern-.025em\copy0\kern-\wd0
  \kern  .05em\copy0\kern-\wd0
  \kern-.025em\raise.0433em\box0 }

\catcode`@=11
\newcount\r@fcount \r@fcount=0
\newcount\r@fcurr
\immediate\newwrite\reffile
\newif\ifr@ffile\r@ffilefalse
\def\w@rnwrite#1{\ifr@ffile\immediate\write\reffile{#1}\fi\message{#1}}

\def\writer@f#1>>{}
\def\referencefile{
  \r@ffiletrue\immediate\openout\reffile=\jobname.ref%
  \def\writer@f##1>>{\ifr@ffile\immediate\write\reffile%
    {\noexpand\refis{##1} = \csname r@fnum##1\endcsname = %
     \expandafter\expandafter\expandafter\strip@t\expandafter%
     \meaning\csname r@ftext\csname r@fnum##1\endcsname\endcsname}\fi}%
  \def\strip@t##1>>{}}

\def\citeall#1{\xdef#1##1{#1{\noexpand\cite{##1}}}}
\def\cite#1{\each@rg\citer@nge{#1}}	

\def\each@rg#1#2{{\let\thecsname=#1\expandafter\first@rg#2,\end,}}
\def\first@rg#1,{\thecsname{#1}\apply@rg}	
\def\apply@rg#1,{\ifx\end#1\let\next=\relax
\else,\thecsname{#1}\let\next=\apply@rg\fi\next}

\def\citer@nge#1{\citedor@nge#1-\end-}	
\def\citer@ngeat#1\end-{#1}
\def\citedor@nge#1-#2-{\ifx\end#2\r@featspace#1 
  \else\citel@@p{#1}{#2}\citer@ngeat\fi}	
\def\citel@@p#1#2{\ifnum#1>#2{\errmessage{Reference range #1-#2\space is bad.}%
    \errhelp{If you cite a series of references by the notation M-N, then M and
    N must be integers, and N must be greater than or equal to M.}}\else%
 {\count0=#1\count1=#2\advance\count1
by1\relax\expandafter\r@fcite\the\count0,%
  \loop\advance\count0 by1\relax
    \ifnum\count0<\count1,\expandafter\r@fcite\the\count0,%
  \repeat}\fi}

\def\r@featspace#1#2 {\r@fcite#1#2,}	
\def\r@fcite#1,{\ifuncit@d{#1}
    \newr@f{#1}%
    \expandafter\gdef\csname r@ftext\number\r@fcount\endcsname%
                     {\message{Reference #1 to be supplied.}%
                      \writer@f#1>>#1 to be supplied.\par}%
 \fi%
 \csname r@fnum#1\endcsname}
\def\ifuncit@d#1{\expandafter\ifx\csname r@fnum#1\endcsname\relax}%
\def\newr@f#1{\global\advance\r@fcount by1%
    \expandafter\xdef\csname r@fnum#1\endcsname{\number\r@fcount}}

\let\r@fis=\refis			
\def\refis#1#2#3\par{\ifuncit@d{#1}
   \newr@f{#1}%
   \w@rnwrite{Reference #1=\number\r@fcount\space is not cited up to now.}\fi%
  \expandafter\gdef\csname r@ftext\csname r@fnum#1\endcsname\endcsname%
  {\writer@f#1>>#2#3\par}}

\def\ignoreuncited{
   \def\refis##1##2##3\par{\ifuncit@d{##1}%
     \else\expandafter\gdef\csname r@ftext\csname
r@fnum##1\endcsname\endcsname%
     {\writer@f##1>>##2##3\par}\fi}}

\def\r@ferr{\endreferences\errmessage{I was expecting to see
\noexpand\endreferences before now;  I have inserted it here.}}
\let\r@ferences=\references
\def\references{\r@ferences\def\endmode{\r@ferr\par\endgroup}}

\let\endr@ferences=\endreferences
\def\endreferences{\r@fcurr=0
  {\loop\ifnum\r@fcurr<\r@fcount
    \advance\r@fcurr by 1\relax\expandafter\r@fis\expandafter{\number\r@fcurr}%
    \csname r@ftext\number\r@fcurr\endcsname%
  \repeat}\gdef\r@ferr{}\endr@ferences}


\let\r@fend=\endpaper\gdef\endpaper{\ifr@ffile
\immediate\write16{Cross References written on []\jobname.REF.}\fi\r@fend}

\catcode`@=12

\citeall\refto		
\citeall\ref		%
\citeall\Ref		%

\def\3he{{$^3${\rm He}}}

\def\eg{{\it e.g.,\ }}

\def\slD{\raise.15ex\hbox{$/$}\kern-.53em\hbox{$D$}}
\def\dsl{\raise.15ex\hbox{$/$}\kern-.57em\hbox{$\Delta$}}
\def\slp{{\raise.15ex\hbox{$/$}\kern-.57em\hbox{$\partial$}}}
\def\nsl{\raise.15ex\hbox{$/$}\kern-.57em\hbox{$\nabla$}}
\def\sla{\raise.15ex\hbox{$/$}\kern-.57em\hbox{$\rightarrow$}}
\def\slla{\raise.15ex\hbox{$/$}\kern-.57em\hbox{$\lambda$}}
\def\slb{\raise.15ex\hbox{$/$}\kern-.57em\hbox{$b$}}
\def\lnp{\raise.15ex\hbox{$/$}\kern-.57em\hbox{$p$}}
\def\lnk{\raise.15ex\hbox{$/$}\kern-.57em\hbox{$k$}}
\def\lnK{\raise.15ex\hbox{$/$}\kern-.57em\hbox{$K$}}
\def\lnq{\raise.15ex\hbox{$/$}\kern-.57em\hbox{$q$}}

\def\cL{{\cal L}}
\def\cM{{\cal M}}

\def\cO{{\cal O}}


\def\pmb#1{\setbox0=\hbox{$#1$}%
\kern-.025em\copy0\kern-\wd0
\kern.05em\copy0\kern-\wd0
\kern-.025em\raise.0433em\box0 }

\def\q2{{Q^2}}
\def\gtwid{\raise.3ex\hbox{$>$\kern-.75em\lower1ex\hbox{$\sim$}}}
\def\ltwid{\raise.3ex\hbox{$<$\kern-.75em\lower1ex\hbox{$\sim$}}}
\def\12{{1\over2}}
\def\part{\partial}

\def\low#1{\lower.5ex\hbox{${}_#1$}}

\def\psl{\raise.15ex\hbox{$/$}\kern-.57em\hbox{$\partial$}}
\def\partt{\raise.15ex\hbox{$\widetilde$}{\kern-.37em\hbox{$\partial$}}}

\def\topppageno1{\global\footline={\hfil}\global\headline
={\ifnum\pageno<\firstpageno{\hfil}\else{\hss\twelverm --\ \folio
\ --\hss}\fi}}

\def\toppageno2{\global\footline={\hfil}\global\headline
={\ifnum\pageno<\firstpageno{\hfil}\else{\rightline{\hfill\hfill
\twelverm \ \folio
\ \hss}}\fi}}

\def\prl#1{Phys.\ Rev.\ Lett.\ {\bf #1}}
\def\prd#1{Phys.\ Rev.\ {\bf D#1}}
\def\plb#1{Phys.\ Lett.\ {\bf #1B}}
\def\npb#1{Nucl.\ Phys.\ {\bf B#1}}

\def\seillac{Nucl.\ Phys.\ (Proc.\ Suppl.) {\bf 4} (1988)}

\def\MeV{{\rm Me\!V}}
\def\GeV{{\rm Ge\!V}}

\def\eg{{\it e.g.},\ }
\def\et{{\it et al.}}

\def\ref#1{${}^{#1}$}
\def\nsection#1 #2{\leftline{\rlap{#1}\indent\relax #2}}

\def\prl#1{Phys.\ Rev.\ Lett.\ {\bf #1}}
\def\prd#1{Phys.\ Rev.\ {\bf D#1}}
\def\plb#1{Phys.\ Lett.\ {\bf #1B}}
\def\npb#1{Nucl.\ Phys.\ {\bf B#1}}

\def\seillac{Nucl.\ Phys.\ {\bf B} (Proc.\ Suppl.) {\bf 4} (1988)}

\def\capri{Nucl.\ Phys.\ {\bf B} (Proc.\ Suppl.) {\bf 17} (1990)}

\def\MeV{{\rm Me\!V}}
\def\GeV{{\rm Ge\!V}}
\def\qa{{quenched approximation}}

{\parindent=0pt
{March 1992
\hfill{Wash. U. HEP/92-60}
}}
\title Chiral Perturbation Theory for the Quenched Approximation of QCD
\author Claude W. Bernard and Maarten F.L. Golterman
\affil Department of Physics
       Washington University
       St. Louis, MO 63130, USA
\abstract
{\parindent=0pt
We describe a technique for constructing the
effective chiral theory for quenched QCD.  The effective
theory which results is a lagrangian one, with
a graded symmetry group which mixes Goldstone bosons and fermions,
and with a definite (though slightly peculiar) set of
Feynman rules.  The straightforward application of these
rules gives automatic cancellation of diagrams which
would arise from virtual quark loops.  The techniques
are used to calculate chiral logarithms in $f_K/f_\pi$, $m_\pi$,
$m_K$, and the ratio of $\langle{\bar s}s\rangle$ to
$\langle{\bar u}u\rangle$.
The leading finite-volume corrections to these
quantities are also computed.
Problems for future study are described.
}
\endtopmatter

\subhead{I.  Introduction and Motivation}

The quenched approximation\refto{quench} to QCD, in which virtual
quark loops are neglected, is a necessary evil
in lattice QCD simulations and will
be with us for the foreseeable future.  Even with the proposed
QCD Teraflop Machine,\refto{tera} the \qa\ will be needed to approach
the crucial corners of the parameter space:  large volumes,
physical quark masses, and the continuum limit.
We therefore need to learn as much as possible analytically
about the \qa\  in order to have good control over the systematics
of such calculations.

In the full theory, chiral perturbation theory (ChPT) is a key
analytic tool.  It gives:
\item{$\bullet$} The detailed form of the approach to the chiral limit.
The universal terms (``chiral logarithms'') can be calculated order
by order in the loop expansion.  Comparison with this expected chiral
behavior provides, for example,  a
crucial check
of lattice weak matrix element calculations.

\item{$\bullet$} The leading finite-volume corrections at
large volume.\refto{finiteV}
As the lightest particles, the pseudoscalar mesons clearly
control these corrections; ChPT is simply the effective theory
of their interactions.

It is therefore clear why one would like to have a ChPT corresponding
to the \qa.  In fact, there have been
several previous attempts
to calculate quenched chiral logarithms.  Morel\refto{morel} and
Sharpe\refto{sharpe1d}
use the strong coupling and $1/d$ expansions;
Kilcup {\it et al.}\refto{kilcup} and Sharpe\refto{sharpecapri}
use the quark-flow approach (see below).
The papers by Sharpe
in particular emphasize the importance
of quenched ChPT and discuss several of the key issues
(in particular, the problems caused by the $\eta'$).
A preliminary version of the current work has been presented in
ref.\ \cite{us}.

\subhead{ II. Quark-flow Approach}

In this approach, one starts with ordinary ChPT for full
QCD and writes down all meson diagrams which contribute
to the process of interest.
To each meson diagram one then associates one
or more quark-flow diagrams in QCD.
Next, one
eliminates all those quark-flow diagrams which have virtual quark
loops.  Finally, one attempts to reinterpret this
elimination as conditions on the meson diagrams.
Note that,  in the
case where more than one quark-flow diagram corresponds
to a given meson diagram, it is
by no means obvious that the final step can always
be performed.  We have been able to carry it through,
more-or-less satisfactorily, in
simple cases (see below), but have been unable to prove --- within
the context of this approach ---
that it can always be done.

In order
to go back and forth between quark flow diagrams
and meson diagrams, the natural basis to use is the
$q\bar q$ basis.  In the neutral sector, this means that one
works with $u\bar u$, $d\bar d$, and $s \bar s$ states
rather than
$\pi^0$, $\eta$, and $\eta'$. The latter
basis is convenient in the full theory since
one can treat the $\eta'$ mass as ``large,'' decouple it,
and work only with $\pi^0$ and $\eta$.
This turns out not to be possible
in the quenched theory.
In full QCD the $\eta'$ gets the singlet part of
its mass ($\equiv\mu$) through the iteration of quark loop diagrams joined
by gluons (see fig.\ 1).

\vbox{
\vskip 4cm
\centerline
{\it Fig.\ 1. Quark flow diagrams for the $\eta'$ propagator in full QCD.}
}

 In the approximation where the $\eta'$ mass
is much greater than the masses of the octet mesons, the $\eta'$
decouples and may be neglected.  In the \qa, on the other hand,
only the first two diagrams in fig.\ 1 survive, and only
the second diagram (the ``two-hairpin'' diagram --- fig.\ 2) depends
on $\mu$.

\vbox{
\vskip 4cm
{\settabs 7 \columns
{\+\hfill\it Fig.\ 2. &\it The ``two-hairpin" diagram, the only diagram which
distinguishes
the singlet\cr
\+ &\it from the octet meson propagator in the \qa.\cr}
}}

The ``two-hairpin vertex'' in fig.\ 2
is $\sim \mu^2$.
Since the vertex is not iterated,
$\mu^2$ appears in the numerator,
not in the denominator, of the
$\eta'$ propagator.  Thus the $\eta'$ cannot be neglected in the
\qa.

In many cases, it is immediately clear which full ChPT diagrams
should be dropped in the \qa.  For example, consider the
lowest order correction to a $\pi^+$ propagator: a meson
tadpole, shown in fig.\ 3.

\vbox{
\vskip 5cm
{\settabs 5 \columns
\+\hfill\it Fig.\ 3. &\it The one-loop contribution to the pion propagator for
full QCD,\cr
\+&\it with
$\pi^+$,
$\pi^0$, $K^+$, $K^0$, $\eta$, and $\eta'$ on the loop.\cr
}}

  When the tadpole is a $K^+$, then the diagram
 must be absent in
the \qa, since the $s$ quark in the $K^+$ is not present
in the external states and must come from a virtual loop.
When the tadpole is itself a $\pi^+$, however, the situation
is less clear. If the $s$ quark of the previous $K^+$ tadpole
is replaced
by a $d$, then again the diagram is absent in the \qa.
But there is now a second possibility: the valence quarks,
themselves, could make the tadpole as in fig.\ 4.

\vbox{
\vskip 5cm
\centerline
{\it Fig.\ 4. A possible
valence quark contribution to the pion
propagator.}
}

The vertex in such a diagram
is a meson-meson scattering vertex
with no quark exchange.
 It turns out that
such a vertex vanishes at ${\cal O}(p^2)$ in ChPT, although
we have never
been able to prove this to  our complete satisfaction within
the quark flow approach.\footnote{*}{
For the number of flavors
$N_F\ge 4$, it is easy to show that this vertex
vanishes.
In that case, one can choose all 4 participating quarks to be
different and thereby make a unique correspondence between the quark
vertex and a meson vertex.  Examination of the trace
structure of the $\cO(p^2)$ chiral lagrangian then immediately
gives the desired result.\refto{steve} However, the proof in this
context for $N_F=3$ escapes
us, though the vertex certainly does vanish,
as can be seen by working backward from
the known result derived in the lagrangian framework of section III.}
  Thus $\pi^+$ tadpoles are also
absent in the \qa.  Indeed, the only correction to the
quenched $\pi^+$ propagator at this order comes from an $\eta'$ tadpole
with a single two-hairpin vertex, fig.\ 5.

\vbox{
\vskip 5cm
{\settabs 4 \columns
\+\hfill\it Fig.\ 5. &\it The quark flow diagram for the one-loop
contribution\cr
\+ & \it to the
pion propagator in quenched QCD.\cr
}}

The end result of this approach can be described as
a ``lagrangian + rules.'' The lagrangian is the ordinary
chiral lagrangian corresponding to full QCD. The rules
give the weighting of the diagrams and prescribe how
to replace $\eta'$ contributions by two-hairpin diagrams.
However, we find this
approach unsatisfactory for two reasons.
First of all, it is difficult to make the application
of the rules routine.  As mentioned above,
it is not obvious that one can always interpret the
elimination of quark-flow diagrams with virtual loops
as conditions on the mesons diagrams. One is, at the minimum,
forced
to prove the vanishing of various vertices (such as
the meson-meson no-exchange vertex).  Not only
are convincing proofs elusive, but new processes
may bring up new such vertices, so it is never clear that all
the necessary proofs have been produced.

A second, more fundamental, problem with this approach is that
the presence of the ``rules'' implies that one does not have a
true lagrangian theory.  This means, for example,
 that the invariance of
the physics under field redefinitions is not guaranteed.
Such redefinitions are needed to reduce
the number of terms in the
lagrangian involving the $\eta'$ ---
see ref. \cite{gasser}.
(These terms are not constrained
much by symmetry because of the anomaly.)
Similarly, the cancellation of the quartic divergences
in ordinary ChPT is guaranteed by the chiral invariance of
the lagrangian and the measure.  It is not clear (at least to us)
whether the rules
automatically respect this cancellation.

We therefore turn now to an alternative approach to quenched
ChPT which gives a true lagrangian framework and makes
the calculation of quenched chiral logarithms routine.

\subhead{III. A Lagrangian Framework}

We start with QCD.  To make a lagrangian which describes
the \qa, we take the ordinary QCD lagrangian and add,
for each quark $q_a$ ($a=u,d,s$), a scalar (ghost) quark
$\tilde q_a$ with the same mass.\refto{morel}
The ghost determinant then cancels
the quark determinant.  Of course, the resulting theory is
not unitary in the quark sector; that is acceptable since the
quenched approximation is not unitary.

Assuming that quark confinement still
holds,  the low-energy effective
theory for this quenched QCD lagrangian may now be constructed.
It will describe the interactions of all
possible pseudoscalar bound states of quarks or scalar quarks
with their antiparticles: ordinary $q\bar q$ mesons ($\pi$, $K$,
$\dots$) which we denote, generically, by $\phi$; ghost $\tilde q
\bar{\tilde q}$ mesons denoted by $\tilde \phi$; and fermionic
mesons $\tilde q \bar q$ and $q \bar{\tilde q}$ denoted by
$\chi$ and $\chi^\dagger$, respectively.

As in ordinary ChPT, the symmetries at the quark level determine
the form of the interactions among the mesons.  The symmetry is
 $U(3|3)_L \times U(3|3)_R$, where $U(3|3)$ is ``almost''
a $U(6)$ among $u,d,s,\tilde u,\tilde d,\tilde s$, but has
a graded structure since
it mixes fermions and bosons.\refto{dewitt}
If we write a matrix $U\in U(3|3)$ in block form as
$$U=\left(\matrix{A&C\cr
                  D&B\cr}\right),\eqno(1)$$
then $A$ and $B$ are $3\times 3$
matrices of commuting numbers; $C$ and $D$, of
anticommuting.
Unitarity is defined as usual: $U^\dagger U = I$.
Hermitian conjugation ($\dagger$) also has the usual definition
(complex conjugation of the usual transpose), but complex conjugation
is defined to switch the order of anticommuting variables:
$(\epsilon_1 \epsilon_2)^* = \epsilon_2^* \epsilon_1^*$.
There is also a cyclic ``supertrace''
defined by $str(U) = tr(A) - tr(B)$, and  a ``superdeterminant,''
$sdet(U) = \exp str \ln U$, with the property $sdet(U_1 U_2)
= sdet(U_1) sdet(U_2)$.          Explicitly,
$$sdet(U)=det(A-CB^{-1}D)/det(B).\eqno(2)$$
Now define the
Hermitian field $\Phi$ and the mass matrix $\cM$ by
$$\Phi\equiv\left(\matrix{\phi&\chi^\dagger\cr
                          \chi&\phi\cr}\right),\;\;\;\;\;\;\;\;\;\;
\cM\equiv\left(\matrix{M&0\cr
                       0&M\cr}\right),\eqno(3)$$
where $$M=
\left(\matrix{m_u&0&0\cr
              0&m_d&0\cr
                  0&0&m_s\cr}\right),\eqno(4)$$
is the usual quark mass matrix.
Note that, to lowest order in $M$, these ChPT quark masses are
the same as those of QCD.


The unitary field
$\Sigma \equiv \exp(2 i \Phi/f)$ transforms as
$\Sigma \rightarrow U_L \Sigma U_R^\dagger$.
The lagrangian invariant under the full $U(3|3)_L \times U(3|3)_R$
is then
$$\cL_{\rm inv}={f^2\over 8}str
(\partial_\mu\Sigma\partial^\mu\Sigma^\dagger)
+v\;str(\cM\Sigma+\cM\Sigma^\dagger),\eqno(5)$$
where $f$ and $v$ are as yet undetermined bare coupling
constants.
This looks very
much like ordinary ChPT.

The anomaly breaks the symmetry group down to
$SU(3|3)_L \times SU(3|3)_R \times U(1)$. The anomalous
field is $\Phi_0 \equiv (\eta' - \tilde \eta')/\sqrt{2}$,
where the minus sign comes from the relative minus sign between
boson and fermion loops.  Under the reduced group,
$\Phi_0 \propto str \ln \Sigma = \ln sdet \Sigma$ is invariant,
so arbitrary functions of $\Phi_0$ can be included in the full
lagrangian, $\cL$.  However, in the current framework
one can redefine $\Sigma$ to simplify $\cL$, much as
in ref.\  \cite{gasser}.  The result is
$$\eqalignno{
&\cL =
V_1(\Phi_0)str
(\partial_\mu\Sigma\partial^\mu\Sigma^\dagger)
+V_2(\Phi_0)str(\cM\Sigma+\cM\Sigma^\dagger)\cr
&\phantom{junk}- V_0(\Phi_0)+V_5(\Phi_0)(\partial_\mu\Phi_0)^2,&(6)\cr
&\cL(\Phi_0=0) \equiv \cL_{\rm inv},&(7)\cr}$$
where the functions $V_i$ can be chosen to be real and even by making use
of the freedom allowed by field redefinitions.  In
ref.\ \cite{gasser} a different choice is made: $V_5$ is set to 0 but
$V_2$ is kept complex.
The potentials $V_3$ and $V_4$
from ref.\ \cite{gasser} are not needed for the purposes of this
paper and have been dropped.  Note that the notation in eqns.\ (6)
and (7) is slightly different from that used in ref.\ \cite{us}.
For the purposes of this paper,
we need only the quadratic terms in $\cL_{\Phi_0}$.
We have
$$\eqalignno{
\cL&= \cL_{\rm inv} + \alpha (\partial_\mu \Phi_0)^2 - \mu^2\Phi_0^2
+ \cdots,\cr
\alpha&\equiv V_5(0),&(8)\cr
\mu^2&\equiv (1/2)V''_0(0).}$$

One can now calculate straightforwardly with $\cL$. Note that
because of the minus sign in the definition of $str$, some
of the fields will have negative metrics.   An unusual
feature occurs in the $\eta',\tilde \eta'$ sector:  terms from
$V_0$ and $V_5$ have a different matrix structure from those
of $\cL_{\rm inv}$,
and one cannot diagonalize the quadratic lagrangian in a momentum
independent way.  This leads us to treat the quadratic terms from
$V_0$ and $V_5$ as vertices.
Iterations of these vertices on the same line then automatically
vanish due to cancellation
between the $\eta'$ and the negative metric $\tilde \eta'$.
This is a manifestation of the fact that the iteration of the
two-hairpin vertex is forbidden in quenched QCD.
When $m_s \not= m$ another peculiarity
occurs: the $\pi^0$ is the only well-behaved neutral particle. The
propagators of the orthogonal states do not have
simple-pole structures.  When also $m_u \not= m_d$, even the
$\pi^0$ propagator becomes ill-behaved.

Because of the unusual structure of the neutral sector,
it is convenient, both in the formalism and in actual
computations,   to write the neutral
meson propagators
in the basis of the states corresponding to $u\bar u$,
$d\bar d$, $s\bar s$ and their ghost counterparts.
 As mentioned above,
this is unlike the case of full QCD, where, due to the singlet part
of the $\eta'$ mass, the propagators in this sector are diagonal in
the $\pi^0$, $\eta$, $\eta'$ basis.

Since we have a true lagrangian theory, the symmetry should guarantee
that any quartic divergences in the diagrammatic expansion
will be cancelled by contributions from the measure, just as in the
full theory.  We have explicitly checked this in the
$SU(1|1)_L \times SU(1|1)_R$  case.  The $SU(3|3)_L \times SU(3|3)_R$ case
is considerably
more complicated;  however, it turns out that no quartic
divergences appear in any of the calculations presented below.

The vertices from $V_0$ and $V_5$  can in principle appear
more than once in a diagram if they occur on different
lines.  However, it is our philosophy to treat the
parameters $\mu^2$  and $\alpha$ as small.  This is certainly
true in the $1/N_c$ expansion.
(Recall that in full QCD the $\eta'$ gets its singlet mass at order
 $1/N_c$.\refto{witten})
Moreover, it appears that the real expansion parameters are
$\alpha/3$ and $\mu^2/3$ (see below).  To estimate the size
of the one-loop corrections, one may take
$\alpha\equiv 0$, neglect $\eta$-$\eta'$ mixing,
and use the physical $\eta'$ mass.  One gets
$\mu^2/3 \simeq (500\ \MeV)^2 \simeq m_K^2$, which
leads one to expect
 that quenched ChPT should be roughly as good as
full ChPT for the kaon.

\subhead{IV. Results and Conclusions}

We have calculated, at one-loop,
$m_\pi$, $m_K$, $f_\pi$, $f_K$, $\langle \bar u u \rangle$, and
$\langle \bar s s \rangle$.  In the isospin limit
($m_u=m_d\equiv m$) and at infinite volume, we get:
$$\eqalignno{
(m_\pi^{\rm 1\!-loop})^2&= m_\pi^2 \left( 1 + {1 \over 8\pi^2f^2}
\left({\alpha \over 3} \Lambda^2 -{\mu^2\over3}
+ {\alpha\over3}m_\pi^2 + \left({\mu^2\over3}
     - {2\alpha\over3}m_\pi^2\right) \ln(\Lambda^2/m_\pi^2)
\right)\right),\cr
(m_K^{\rm 1\!-loop})^2&= m_K^2\Biggl(1 + {1 \over 8\pi^2f^2}
\Biggl({\alpha \over 3} \Lambda^2 + \left({\mu^2\over3}
     - {2\alpha\over3}m_K^2\right) \ln(\Lambda^2/m_\pi^2)\cr
    &\hphantom{=\;}
 - \left({\mu^2\over 3}-{\alpha\over 3}(2m_K^2-m_\pi^2)\right)
     {{(2m_K^2-m_\pi^2)}\over{2(m_K^2-m_\pi^2)}}
    \ln\left({2m_K^2 \over m_\pi^2} - 1 \right)\Biggr)\Biggr),\cr
f_\pi^{\rm 1\!-loop}&= f,&(9)\cr
\left({f_K\over f_{\pi}}\right)^{\rm 1\!-loop} &= 1 +
{1 \over 16\pi^2 f^2} \left(
                   {\alpha\over3} m_K^2 -{\mu^2\over 3} +
{{\mu^2\over 3} m_K^2\! - \!
{\alpha\over 3} m_\pi^2\left(2m_K^2-m_\pi^2\right) \over 2\left(m_K^2 -
                  m_\pi^2\right)} \ln\left({2m_K^2 \over m_\pi^2}
- 1 \right) \right),\cr
{m\langle{\bar u}u\rangle}^{\rm 1\!-loop}
&=-{1\over 4}(m_\pi^{\rm 1\!-loop})^2f^2\cr
{m_s\langle{\bar s}s\rangle}^{\rm 1\!-loop}
&=-{1\over 4}
(2m_K^2-m_\pi^2)f^2\Biggl(1+ {1 \over 8\pi^2f^2}
\Biggl({\alpha\over 3}\Lambda^2+\cr
&\hphantom{=\;}
\Biggl({\mu^2\over3}-{{2\alpha}\over 3}(2m_K^2-m_\pi^2)\Biggr)
\ln\left(\Lambda^2/(2m_K^2-m_\pi^2)\right)
+{\alpha\over 3}(2m_K^2-m_\pi^2)
-{{\mu^2}\over 3}
\Biggr)\Biggr),
}$$
where $\Lambda$ is the cutoff, and
$m_K$, $m_\pi$, and $f$ are the bare parameters:
$$m_\pi^2 = {8 v m \over f^2}\ ,\qquad m_K^2 = {4 v (m_s +m) \over f^2}
\ .\eqno(10)$$
  It should be noticed
that, except for the $\Lambda^2$ terms, $\alpha$ terms are actually higher
order in a combined expansion in $1/N_c$ and $M$.  This implies that,
apart from quadratically divergent terms, we may set $\alpha=0$ in
eq.\ (9) systematically.

Note that in the \qa\ the ratio $f_K/f_\pi$ is finite at one loop, unlike
the full theory where this quantity contains a logarithmic divergence.
In fact, if we consider ratios in which the quadratic divergence cancels,
and then set $\alpha=0$ as argued above, the ratios
$(m_K^{\rm 1\!-loop}/m_\pi^{\rm 1\!-loop})^2$ and
$({\langle{\bar s}s\rangle}/\langle{\bar u}u\rangle)^{\rm 1-loop}$ are
also finite.  Expressed in terms of the bare quark masses we have
$$\eqalign{
\left({{m_K^{\rm 1\!-loop}}\over{m_\pi^{\rm 1\!-loop}}}\right)^2&=
{{m+m_s}\over{2m}}\Biggl(1 + {{\mu^2/3} \over 8\pi^2f^2}\Biggl(
1 - {{m_s}\over{m_s-m}}\ln(m_s/m)\Biggr)\Biggr),\cr
{{\langle{\bar s}s\rangle^{\rm 1-loop}}\over{\langle{\bar u}u\rangle^{\rm
1-loop}}}&=
1 - {{\mu^2/3} \over 8\pi^2f^2}\ln(m_s/m).}
\eqno(11)$$

Despite the fact that many of our one-loop
results are finite, they are not quantitative predictions
because the terms in the $\cO(p^4)$ lagrangian
(Gasser and Leutwyler's\refto{gasser} $L_i$'s) may also
contribute.  In other words, we have computed the ``chiral
logarithms'' only, not what are usually called the ``finite terms''
(which are always uncomputable in ChPT).
One would need in general to take further ratios of
physical quantities
to eliminate such uncertainties.

Using $\alpha=0$ and $\mu$ as
estimated above and neglecting the ``finite terms,''
 $\left(f_K/f_{\pi}\right)^{\rm 1\!-loop} \cong
1.07$, indicating that quenched ChPT is working well.
Note however that $(m_\pi^{\rm 1\!-loop}/m_\pi)^2
\cong 1.5$ for $\Lambda\cong 1\GeV$ and
$({\langle{\bar s}s\rangle}/\langle{\bar u}u\rangle)^{\rm 1-loop} \cong 0.4$,
although these
ratios are not directly physical,
and the large corrections should perhaps not be worrisome.

We have also
computed the leading finite volume corrections to the above results.
The calculation is straightforward: we simply replace the
infinite-volume meson propagators by their finite-volume counterparts.
We find:
$$\eqalign{
\Delta\left((m_\pi^{\rm 1-loop})^2\right)&=
{{m_\pi^2}\over{4\pi^2f^2}}(\mu^2-\alpha m_\pi^2)
\sqrt{{2\pi}\over{m_\pi L}}e^{-m_\pi L},
\cr
\Delta\left((m_K^{\rm 1-loop})^2\right)&=0,
\cr
\Delta\left(f_\pi^{\rm 1-loop}\right)&=0,
\cr
\Delta\left((f_K/f_\pi)^{\rm 1-loop}\right)&=
{1\over{16\pi^2f^2}}(\mu^2-\alpha m_\pi^2)
\sqrt{{2\pi}\over{m_\pi L}}e^{-m_\pi L},
\cr
\Delta\left(m\langle {\bar u} u\rangle^{\rm 1-loop}\right)&=
-{{m_\pi^2}\over{16\pi^2}}(\mu^2-\alpha m_\pi^2)
\sqrt{{2\pi}\over{m_\pi L}}e^{-m_\pi L},
\cr
\Delta\left(m_s\langle {\bar s} s\rangle^{\rm 1-loop}\right)&=0,
\cr
}\eqno(12)$$
where $\Delta({\rm quantity})$ denotes leading
order finite volume corrections to be added
to the infinite volume one-loop expressions for the quantities given
in eq.\ (9).
$L$ is the spatial size of the box.  We  have assumed
periodic boundary conditions\footnote{*}{In order
to preserve the graded symmetry, the ghost quarks
must have the same boundary conditions (antiperiodic or periodic)
as are chosen for the quarks. All mesons, including
the fermionic ones, will thus have periodic boundary conditions.}
   and $T\gg L$, where $T$  is the
temporal size of the box.
In addition we have neglected terms of order $e^{-m_\pi(\sqrt{2}L)}$
and of order $e^{-m_K L}$. Thus the leading corrections
come only from pions propagating from the closest periodic
images of the original box.   Finally, we have taken
only the leading contribution to  the pion progagator and
neglected terms of order $(m_\pi L)^{-{3\over2}}e^{-m_\pi L}$.
Note that the terms we have neglected may
very well not be small in many current lattice
simulations.  In such cases,
the exact one-loop finite volume corrections (computed by
using the exact finite-volume propagators) should be used.

Some comments on our results and directions for future
work:

\item{$\bullet$} The absence of chiral logarithms in $m_\pi$ seen
in ref. \cite{sharpe1d} is presumably a feature of the
leading term in the $1/d$ expansion.  Indeed,
the $\eta'$ diagrams which give such logarithms are mentioned
in ref. \cite{sharpecapri}.

\item{$\bullet$} Many of these results
(\eg $\left(f_K/f_{\pi}\right)^{\rm 1\!-loop}$,
$({\langle{\bar s}s\rangle}/\langle{\bar u}u\rangle)^{\rm 1-loop}$,
or $m_\pi^{\rm 1-loop}/ m_\pi$)
blow up as $m\to 0$. This is
an IR effect coming from the double pole of the two-hairpin
diagram and is absent in the full theory where the vertex
is iterated.  It is not clear at this point whether
this is a sickness of the \qa\  or only of the current
quenched chiral expansion.

\item{$\bullet$} A Gasser-Leutwyler\refto{gasser}
program for quenched ChPT at one-loop is possible:
we expect there to be
interesting numerical relations involving only
computable (on the lattice) quantities.  Such relations could give
quantitative insight into the effects of quenching.

\item{$\bullet$}  The techniques described here can be easily
used to generate the effective theory for a QCD in which the
quark loops are not neglected, but the masses of valence
and virtual quarks are not identical.  Such an effective theory
is relevant to many
``full QCD'' simulations.  Similarly, one can generate effective
theories which correspond to the quenching of some, but not all
the light quarks.

\item{$\bullet$} It seems to be
straightforward to extend these ideas to
the calculation of the chiral logarithms in weak matrix elements.

\subhead{Acknowledgements}

We are grateful to
Steve Sharpe and Jan Smit for useful discussions. We thank the
Department of Energy
 for partial support under grant number DE-AC02-78ER04915.
This work was begun at the Institute for Theoretical Physics,
Santa Barbara,
and was partially supported
under NSF grant number PHY89-04035.  Additional work was done at UCLA
under  NSF grant number PHY89-15286.

\references

\refis{steve}
S. Sharpe, private communication.

\refis{us} C.\ Bernard and M.\ Golterman,
Wash.\ U.\ HEP/91-31,
     to be published in the proceedings of the 1991 Conference on Lattice
     Field Theory, KEK, Tsukuba, Japan.

\refis{quench} H.\ Hamber and G.\ Parisi, \prl{47} (1981) 1792;
E.\ Marinari, G.\ Parisi and C.\ Rebbi, \prl{47} (1981)
1795; D.\ H.\ Weingarten, \plb{109} (1982) 57.

\refis{tera}
QCD Teraflop Collaboration (S.~Aoki {\it et al.}),
Intl.\ J.\ Mod.\ Phys.\ {\bf C2} (1991) 829.

\refis{finiteV}
J. Gasser and H. Leutwyler, \plb{184} (1987) 83,
\plb{188} (1987) 477,
and \npb{307} (1988) 763; H. Leutwyler, \seillac\ 248
and \plb{189} (1987) 197; H.\ Neuberger, \npb{300} (1988) 180;
P.\ Hasenfratz and
H. Leutwyler, \npb{343} (1990) 241.

\refis{morel}
A.\ Morel, J.\ Physique {\bf 48} (1987) 111.

\refis{sharpe1d}
S.\ Sharpe, \prd{41} (1990) 3233.

\refis{kilcup}
G. Kilcup \et, \prl{64} (1990) 25.

\refis{sharpecapri}
S.\ Sharpe, \capri\ 146 and DOE/ER/40614-5, to be published in
{\it Standard Model, Hadron Phenomenology and Weak Decays on the
Lattice},
ed.\ G.\ Martinelli, World Scientific.

\refis{gasser}
J.\ Gasser and H.\ Leutwyler, \npb{250} (1985) 465.

\refis{dewitt} For a description of the
properties of graded groups, see for example, B.\ DeWitt,
{\it Supermanifolds}, Cambridge, 1984.

\refis{witten}
E.\ Witten, \npb{156} (1979) 269;
G.\ Veneziano, \npb{159} (1979) 213.

\endreferences

\endit